# Linearization of the response of a 91-actuator magnetic liquid deformable mirror

Denis Brousseau,[1,*] Ermanno F. Borra,[1] Maxime Rochette,[1] and Daniel Bouffard Landry[1]

[1]*Département de physique, de génie physique et d'optique and Centre d'optique, photonique et laser (COPL), Université Laval, 2375 rue de la Terrasse, Québec, Québec, CANADA G1V 0A6*
[*]*denis.brousseau@copl.ulaval.ca*

**Abstract:** We present the experimental performance of a 91-actuator deformable mirror made of a magnetic liquid (ferrofluid) using a new technique that linearizes the response of the mirror by superposing a uniform magnetic field to the one produced by the actuators. We demonstrate linear driving of the mirror using influence functions, measured with a Fizeau interferometer, by producing the first 36 Zernikes polynomials. Based on our measurements, we predict achievable mean PV wavefront amplitudes of up to 30 μm having RMS residuals of λ/10 at 632.8 nm. Linear combination of Zernikes and over-time repeatability are also demonstrated.

## References and links

1. J. Liang, D. R. Williams, and D. T. Miller, "Supernormal Vision and High-Resolution Retinal Imaging Through Adaptive Optics," J. Opt. Soc. Am. A **14,** 2884–2892 (1997).
2. R. El-Agmy, H. Bulte, A. H. Greenaway, and D. Reid, "Adaptive beam profile control using a simulated annealing algorithm," Opt. Express **13**, 6085-6091 (2005).
3. M. Ogasawara and M. Sato, "The applications of a liquid crystal aberration compensator for the optical disc systems," in *Adaptive Optics for Industry and Medicine*, Ed. J C Dainty, Imperial College Press, London, 369-375 (2008).
4. P. Laird, E. F. Borra, R. Bergamesco, J. Gingras, L. Truong, and A. Ritcey, "Deformable mirrors based on magnetic liquids," Proc. SPIE **5490**, 1493-1501 (2004).
5. E. F. Borra, A. M. Ritcey, R. Bergamasco, P. Laird, J. Gingras, M. Dallaire, L. Da Silva, and H. Yockell-Lelievre "Nanoengineered Astronomical Optics," Astron. Astrophys. **419**, 777-782 (2004).
6. G. Vdovin, "Closed-loop adaptive optical system with a liquid mirror," Opt. Lett. **34**, 524-526 (2009).
7. D. Brousseau, E. F. Borra, and S. Thibault, "Wavefront correction with a 37-actuator ferrofluid deformable mirror," Opt. Express **15**, 18190-18199 (2007).
8. D. Brousseau, E. F. Borra, S. Thibault, A. M. Ritcey, J. Parent, O. Seddiki, J.-P. Dery, L. Faucher, J. Vassallo, and A. Naderian, "Wavefront correction with a ferrofluid deformable mirror: experimental results and recent developments," Proc. SPIE **7015**, 70153J (2008).
9. A. Iqbal, A. and F. B. Amara, "Modeling of a Magnetic-Fluid Deformable Mirror for Retinal Imaging Adaptive Optics Systems," International Journal of Optomechatronics **1**, 180-208 (2007).
10. A. Iqbal, A. and F. B. Amara, "Modeling and experimental evaluation of a circular magnetic-fluid deformable mirror," International Journal of Optomechatronics **2**, 126-143 (2008).
11. J. Parent, E. F. Borra, D. Brousseau, A. M. Ritcey, J.-P. Déry, and S. Thibault, "Dynamic response of ferrofluidic deformable mirrors," Appl. Opt. 48, 1-6 (2009).
12. R. S. Caprari, "Optimal current loop systems for producing uniform magnetic fields," Meas. Sci. Technol. **6**, 593-597 (1995).
13. K. E. Moore and G. N. Lawrence, "Zonal model of an adaptive mirror," Appl. Opt. **29**, 4622-4628 (1990).
14. J. Alda and G. D. Boreman, "Zernike-based matrix model of deformable mirrors: optimization of aperture size," Appl. Opt. **32**, 2431-2438 (1993).
15. M. M.-Hernandez, M. Servin, D. M.-Hernandez, G. Paez, "Wavefront fitting using Gaussian functions," Optics Communications **163**, 259-269 (1999).
16. L. Thibos, R. A. Applegate, J. T. Schweigerling, and R. Webb, "Standards for reporting the optical aberrations of eyes," in *OSA Trends in Optics and Photonics* **35**, 232-244 (2000).

17. S. Thibault, 2006 Feb. 14 "Method and System for Characterizing Aspheric Surfaces of Optical Elements." United States Patent US 6,999,182.

## 1. Introduction

Deformable mirrors were originally developed for Astronomy but they are now used in other applications such as ophthalmologic, laser beam shaping and industrial applications [1-3]. Most deformable mirrors presently available use solid thin plates or membranes. Building deformable mirrors having large number of actuators is very expensive and they are typically limited to strokes of only a few microns. Although MEMS deformable mirrors demonstrated potential for low-cost and high number of actuators, they are still limited in available stroke. MEMS deformable mirrors are also not appropriate for optical testing applications where large mirror diameters are often required. A new promising technology to build a liquid deformable mirror that uses magnetic liquids (ferrofluids) has been suggested by Borra et al. in [4]. Ferrofluidic deformable mirrors (FDMs) have the major advantage over solid ones that these liquids have extremely smooth surfaces that naturally follow the equipotential surfaces created by magnetic fields. FDMs can have smooth deviations from flatness that can be as small as a few nanometers to as large as several millimeters. The other advantage that FDMs have with respect to solid deformable mirrors is their low cost per actuator.

Ferrofluids have low reflectance and must be coated with a reflective layer. Although it was not necessary for this paper, we usually coat them with MeLLFs (Metal Liquid-Like Films) [5]. We also began working on coating ferrofluids with reflective elastomeric membrane coatings. FDMs are not the only approach that has been considered to make liquid deformable mirrors. Another approach using electrostatically deformed liquids is also presented in [6].

Many technical improvements have been made since Borra et al. in [7, 8]. However, there remained a major inconvenience with the early-generation of FDMs that came from the fact that the surface of the liquid is shaped by a magnetic vector field and that the induced deformations depends on the square of the magnetic field, requiring novel complicated control algorithms [7]. A major technical breakthrough that overcomes these problems was recently proposed by Iqbal and Amara [9, 10]. The simple, yet powerful, technique described in [9, 10] superposes a strong and uniform magnetic field to the magnetic field of the actuators, thereby linearizing the response of the FDM. The major advantage of this linearization is that one can use the same proven algorithms that are used with solid deformable mirrors.

Until recently, FDMs were thought to be restricted to corrections at frequencies lower than 10 Hz, thus limiting their usefulness. Recent experiments by Parent et al. [11] demonstrate that the operational frequency of FDMs can be increased to 1 kHz by increasing the viscosity of ferrofluids.

In this article, we discuss wavefronts measurements produced by a 91-actuator FDM that uses this linearization technique. The experiments show high surface qualities and large strokes, demonstrating that FDMs are credible competitors to solid deformable mirrors

## 2. Theory

*2.1 Linearization of the response of FDMs*

The deformation $h$ produced on the ferrofluid is proportional to the square of the total magnetic field at the ferrofluid-air interface [7]. The technique to linearize the behavior of FDMs presented in [9, 10] consists of superposing a strong and uniform magnetic field to that produced by the actuators. The deformation $h$ on the liquid surface caused by one actuator is then given by,

$$h = kB^2 = k(b+B_0)^2 = k\left(b^2 + 2bB_0 + B_0^2\right), \tag{1}$$

where $b$ is the magnetic field of the actuator, $B_0$ the external and uniform magnetic field with $B_0 \gg b$, and $k$ is a constant that depends on the physical properties of the liquid. The magnetic field $B_0$ being uniform and $b$ very small compared to $B_0$, the only term that contributes to the local deformation is the term $2bB_0$. This has the effect of linearizing the response of the actuators ($h$ directly proportional to $b$) and also amplifies the maximum amplitude the mirror can produce ($h$ directly proportional to $B_0$). This method was demonstrated experimentally by Iqbal and Amara [9, 10] and later confirmed by Brousseau et al. [8].

Iqbal and Amara [10] used a Helmholtz coil in their setup to produce the required uniform external magnetic field. A Helmholtz coil produces a uniform magnetic field up to the 4$^{th}$ order derivative with respect to the position near the center of the coil [12]. While Iqbal and Amara were able to successfully demonstrate the linearization technique using a Helmholtz coil, they state, though they do not quantify, that the non-uniformity produced a curvature on the initial surface of the liquid [10]. Even if this residual wavefront can be compensated using the actuators, this is done at the expense of using stroke that would otherwise be available for the correction of incoming wavefronts. In regard of this, we opted to use a Maxwell coil instead of a Helmholtz coil. A Maxwell coil is made from 3 separate coils (see Fig.1) and produces a uniform magnetic field up to the 6$^{th}$ order derivative with respect to the position near the center of the assembly [12]. The radii of the coils and their vertical position must respect the values given in Fig. 1. The number of ampere-turns of both the lower and upper coils must be exactly in the ratio of 49/64 relative to the middle coil. For design simplicity, it is much easier to adjust the number of turns of the lower and upper coils than to adjust their current. This way, the three coils can be arranged in a series circuit and supplied with a single current value.

When the ratio of turns of the lower and upper coils is exactly 49/64 with respect to the middle one, the magnetic field at the center of a Maxwell coil is given by

$$B_0 = \frac{15}{16} \mu_0 \frac{NI}{R}, \qquad (2)$$

where $N$ is the number of turns of the middle coil, $R$ is the radius of the middle coil and $I$ is the current supplied to the Maxwell coil.

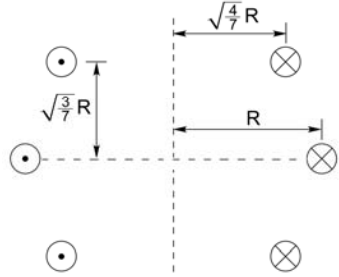

Fig. 1. Schematic of a Maxwell coil. The number of ampere-turns of the lower and upper coils must be exactly in the ratio 49/64 relative to the middle coil. Adjusting the number of turns following this proportion, instead of adjusting the current for each coil, represents the best choice as it allows assembling the three coils in a series circuit.

## 2.2 Control of the FDM

The linear driving method of FDMs allows the use standard methods of controlling the FDM surface. A wavefront $\mathbf{w}_m$ produced by the deformable mirror is then given by a linear combination of the individual response functions of the actuators in the matrix form [13]:

$$\mathbf{w}_m = \mathbf{Ha}, \qquad (3)$$

where **H** is called the influence matrix and **a** is a vector made of the control signals of the actuators. For the case of a FDM, the control signals are the currents supplied to the actuators. Each column of the control matrix **H** represents the response function of a single actuator shifted to its corresponding location, while each row of **H** corresponds to a single wavefront data sample. The matrix **H** is usually rectangular and needs to have more rows (data samples) than columns (actuators). To obtain a given targeted wavefront **w**, the solution for minimum variance of Eq. 3 gives the following vector of control signals to supply to the deformable mirror [12]:

$$\mathbf{a} = (\mathbf{H}^t \mathbf{H})^{-1} \mathbf{H}^t \mathbf{w}, \quad (4)$$

where the superscript **t** denotes the transpose matrix operation. Assuming that the mean wavefront error is zero, the total squared error between the targeted wavefront **w** and the wavefront $\mathbf{w}_m$ produced by the deformable mirror is:

$$\sigma^2 = \boldsymbol{\varepsilon}^t \boldsymbol{\varepsilon}, \quad (5)$$

where **ε** is the wavefront error:

$$\boldsymbol{\varepsilon} = \mathbf{w} - \mathbf{w}_m. \quad (6)$$

The RMS residual wavefront error is obtained by taking the square root of the total squared error divided by the number of data samples $N$ and is given by

$$\sigma_{RMS} = \sqrt{\frac{\boldsymbol{\varepsilon}^t \boldsymbol{\varepsilon}}{N}}. \quad (7)$$

## 3. The 91-actuator deformable mirror

The FDM consists of 91 2.8-mm diameter custom coils (actuators) made by Dia-Netics, Inc. and hexagonally arranged (see left of Fig. 2). Each resin-coated coil consists of about 300 turns of AWG36 magnet wire wound on a brass core having a 1-mm diameter and has a total length of 15 mm. The resistance of each coil is 2.8 ohms. The actuators are supplied in current by a custom amplifying stage that can deliver -200 to +200 mA to each actuator at a resolution of 6 µA. The amplifying stage is controlled by a PD2-AO-96-16 96/16-bit analog output channels PCI card from United Electronic Industries. The current in each actuator is controlled using LabVIEW.

The Maxwell coil was made by our machine shop from an aluminum piece and has a total height of 120 mm (see right of Fig. 2). The form was sent for winding with AWG20 magnet wire following the turn ratio shown in Fig. 1. The total resistance of the device is 7.5 ohms. Current in the Maxwell coil is supplied by a 0-10 A stabilized power supply. Using Eq. 2, the magnetic field produced at the center of the Maxwell coil is about 40 gauss when supplied with a 1.0 A current.

The FDM is placed within the Maxwell coil with the top portion of the actuators lying near the middle of the coil form, where the magnetic field is uniform (see Fig. 3). A container filled with a 1-mm depth of EFH1 ferrofluid from Ferrotec Corp. sits on top of the actuators. A circular 50-mm diameter optical-quality BK7 window is used to protect the liquid surface from dust particles and air currents from the room air exchange system.

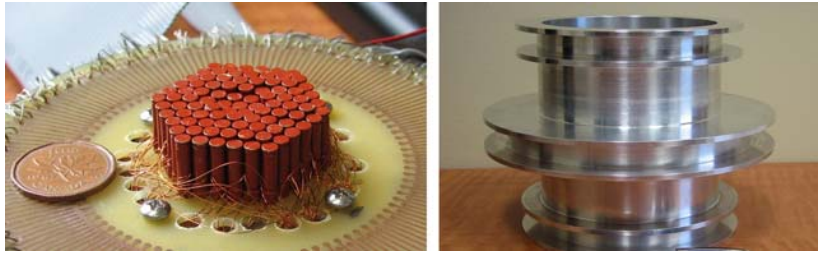

Fig. 2. Pictures of the 91-actuator FDM (left) and Maxwell coil (right). The Maxwell coil is shown prior to winding. The three winding areas of Fig. 1 can clearly be seen.

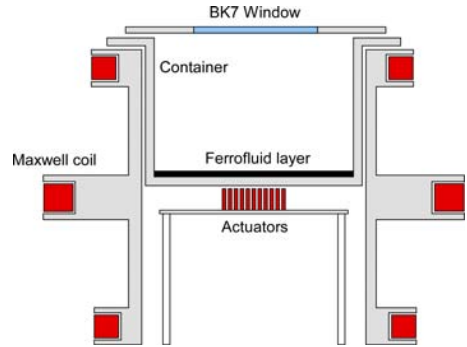

Fig.3. Schematic of the FDM assembly. The total height is 120 mm and the inside diameter of the container is 80 mm. A 1-mm thick layer of EFH1 ferrofluid is used.

## 4. Results

Our wavefront measurements were carried out using two different instruments: A Fizeau interferometer made by the ZYGO Corporation and a HASO HR44 Shack-Hartmann wavefront sensor made by Imagine Optics. The ZYGO interferometer was used for most of the measurements since it gives an excellent spatial resolution of the mirror surface and only requires a fold mirror for the measurements. All surface amplitudes given by the ZYGO were scaled by a factor of two to give wavefront amplitudes. However, the ZYGO interferometer cannot measure wavefronts having large amplitudes. While the HASO sensor has a poorer spatial resolution, it can measure large amplitude wavefront deformations. Consequently, we used it to make linearity measurements and produce a wavefront example at high amplitudes. All wavefront data from the HASO wavefront sensor were reconstructed using Imagine Optics software in zonal reconstruction mode.

*4.1 Driving the FDM*

A current of 500 mA was supplied to the Maxwell coil to produce a constant magnetic field of about 20 gauss at the liquid-air interface. With the Maxwell coil in operation, we measured an initial residual wavefront having a RMS amplitude of 0.10 μm compared to a RMS wavefront amplitude of 0.07 μm when the Maxwell coil was not in operation. Each actuator was then successively supplied with a current of 20 mA and its influence function recorded using the ZYGO MetroPro software. The pupil size of 23 mm in diameter was chosen to exclude the outer ring of actuators so as to better reproduce wavefronts having large slopes at the edge of the pupil [15]. Though the outer ring of actuators was excluded from the pupil, those actuators were active when using the FDM. The ratio of the pupil size relative to the full size diameter of the mirror is shown in Fig. 4.

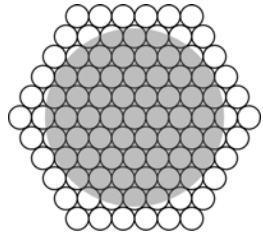

Fig.4. Diameter of the pupil (shaded area) relative to the full diameter of the mirror used in for the ZYGO measurements. Pupil size is 23 mm and full diameter of the actuator array is 33 mm. The outer ring of actuators is kept active when using the FDM.

Fig. 5 shows the typical *x* and *y* profiles of an actuator influence function recorded using the HASO wavefront sensor. Apart from slight differences in amplitudes, all actuators, including the ones at the pupil edge, produce identical influence function profiles. This is easily explained as there is any physical constraint on the mirror surface at the edges of the actuators region. As in [7], the influence function is well described by a Gaussian profile and corresponds to a coupling constant of about 45%. Based on the response of a single actuator, we deduced a maximum stroke value of 8.75 μm for a single actuator at the maximum operating current of the stage amplifier. The stroke can easily be doubled by increasing the Maxwell coil current from 0.5 to 1.0 A and even higher strokes could be achieved using higher Maxwell coil currents.

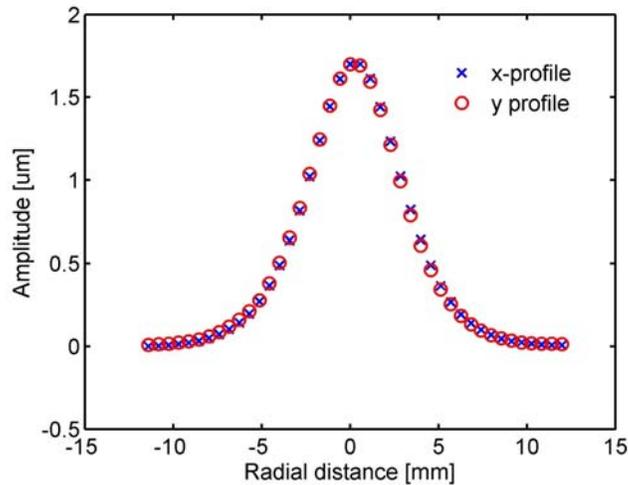

Fig. 5. Typical *x* and *y* profiles of the influence function of a single actuator when supplied with a current of 20 mA. The wavefront profiles were obtained using the HASO wavefront sensor.

Fig. 6 shows an example of the linearity advantage obtained using the new driving technique. The data points in blue on the figure shows, the influence functions of the central actuator (#1) and its nearest left neighbor (#2) when both are independently driven in opposite directions (push and pull), along with the response when both are driven simultaneously (red dashed line), and the arithmetic addition of the individual response of each actuator (black dashed line). Fig. 6 also illustrates that the actuators now have push-pull ability as opposed to driving FDMs without an external magnetic field.

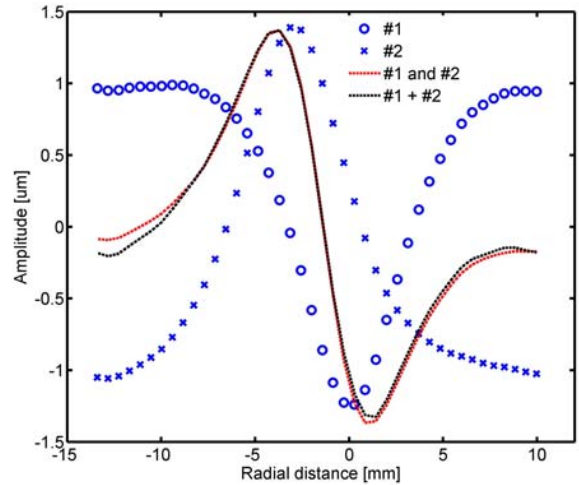

Fig. 6. Linear addition of influence functions. Data shown in blue are for actuator #1 (at pupil center) and actuator #2 (nearest left neighbor) when both are actuated separately in opposite directions. The red dashed line shows the response of the FDM when both actuators are driven simultaneously. The black dashed curve shows the arithmetic addition of the individual responses. The wavefront profiles were obtained using the HASO wavefront sensor.

Fig. 7a shows the PV amplitude of the influence function of the FDM central actuator as a function of current when the Maxwell coil is supplied with a constant current of 0.5 A. A linear fit of the data (in red) shows that the response of the actuator has become linear. Fig. 7b shows the PV amplitude of the influence function of the FDM central actuator when it is supplied with a constant current of 20 mA, but current in the Maxwell coil is varied from 0.2 to 1.0 A. Again, the response is linear and confirms that the Maxwell coil can be used to both linearize and amplify the response of FDMs.

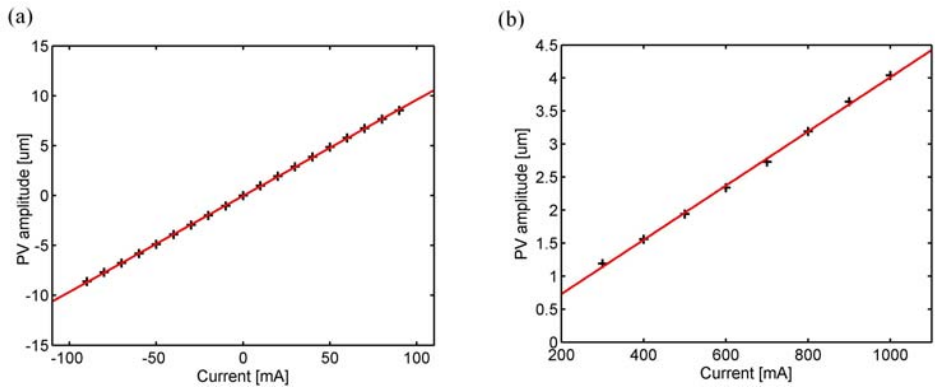

Fig. 7. (a) Amplitude response of the FDM central actuator as a function of current when the Maxwell coil is supplied with a constant current of 0.5 A. (b) Amplitude response of the FDM central actuator when the current in the Maxwell coil is varied from 0.2 to 1.0 A. A current of 20 mA in the central actuator was used for (b). Linear fits of the data are shown in red. The wavefront amplitudes were obtained using the HASO wavefront sensor.

The control matrix **H** was constructed from the recorded influence functions of the 91 actuators. The required signals (currents) to produce the first 36 Zernike polynomials following the OSA numbering scheme [16] were computed from Eq. 4 using singular value decomposition. Each Zernike polynomial was targeted to a PV wavefront amplitude of 4 μm. The wavefront error $\varepsilon$ was calculated for each Zernike using Eq. 6. Currents to reproduce this residual wavefront error were then computed using Eq. 4. The Zernikes were then further optimized by subtracting this current vector from the currents of the first run using a gain of 0.6. A third iteration on the currents was found to be unnecessary since the residuals stopped to decrease. Fig. 8 shows Zernikes $Z_2^0$ to $Z_4^0$ after the second current iteration. In principle, these iterations should only be done once for each term. Once the best current vector of each Zernike is found and unless the ferrofluid is replaced, it should be possible to keep these currents as a reference set to reproduce Zernikes of different amplitudes by scaling their appropriate current vector. Measurements of a scaled defocus appearing further down in this section confirm this (see Fig. 10). Section 4.3 also confirms this by showing the over-time repeatability of some Zernikes using the optimized current vectors.

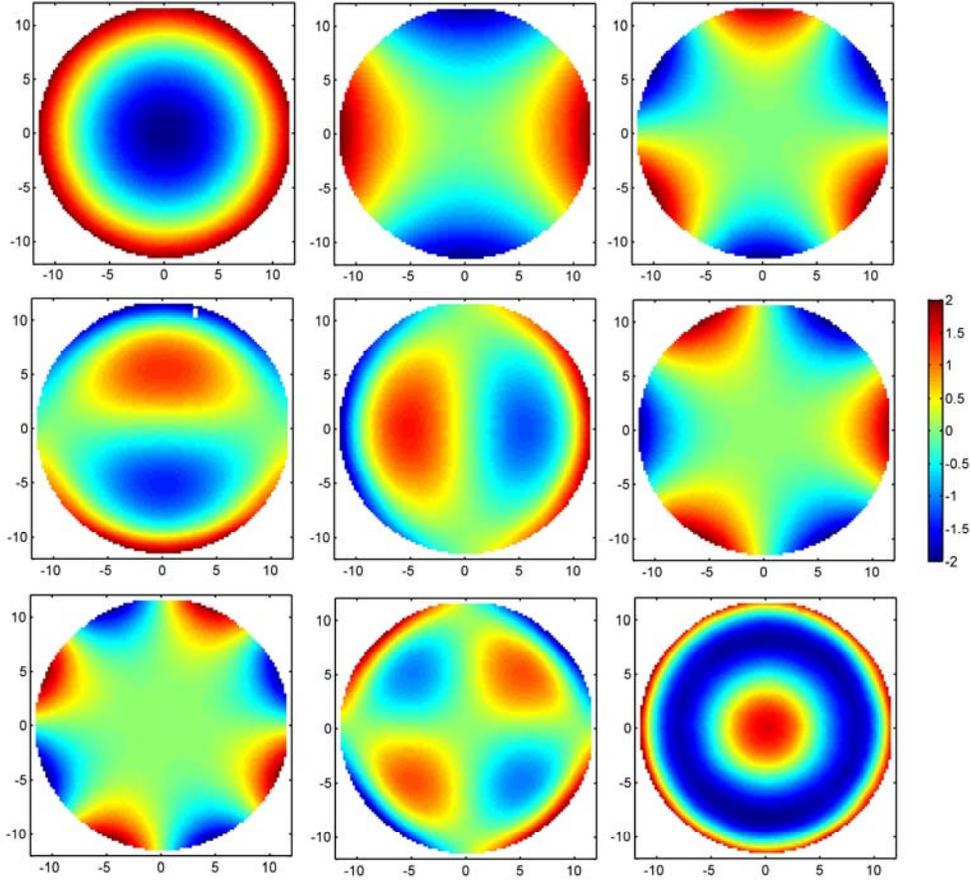

Fig. 8. Zernike polynomials from $Z_2^0$ to $Z_4^0$ produced by the FDM and measured with a ZYGO interferometer. Grid units are in millimeters and wavefront amplitude is measured in µm.

The RMS wavefront residuals were calculated for each Zernike using Eq. 7 and results appear as a bar chart in Fig. 9a. The higher residual RMS errors seen on terms 12 and 24 ($Z_4^0$ and $Z_6^0$) are in agreement with numerical simulations that use the addition of Gaussian influence functions over a circular pupil. Based on those residual errors, we extrapolated the maximum wavefront amplitude coefficient that would be achievable with the 91-actuator FDM while keeping constant the residual wavefront RMS error to λ/10 at 632.8 nm. The results are given by the bar chart in Fig. 9b. It can be seen that the achievable mean coefficient amplitude over all terms is over 15 µm with some terms even over 35 µm while still keeping a residual error of λ/10. Note that wavefronts having even higher amplitudes and comparable residuals would be possible by increasing the number of actuators and using a higher current in the Maxwell coil.

As an example of the performance of the FDM, we corrected the initial wavefront caused by the Maxwell coil on the liquid surface. The initial surface of the mirror was successfully flattened from a wavefront RMS amplitude of 0.10 down to a wavefront RMS value of 0.006 µm.

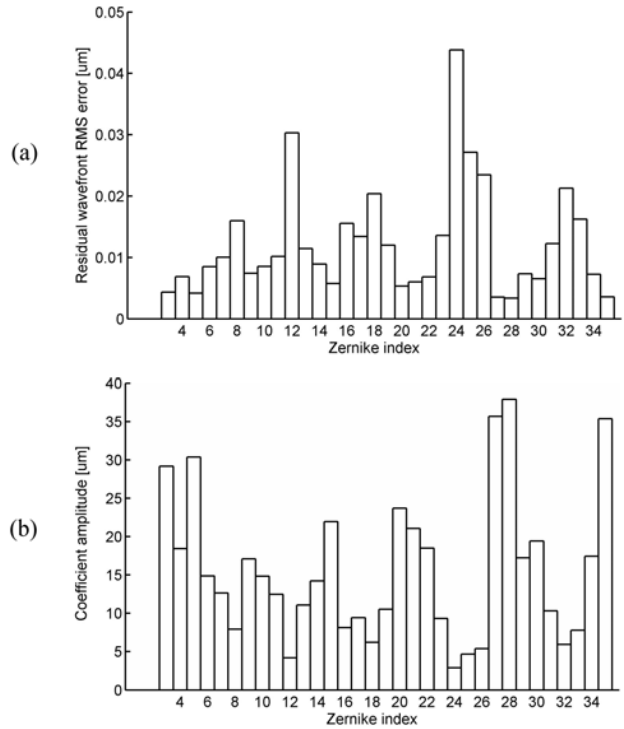

Fig. 9. (a) Residual wavefront RMS error for the first 36 Zernike polynomials (excluding piston, tip and tilt) obtained with the FDM. Targeted PV wavefront amplitude of the Zernikes was 4 µm. (b) Computed maximum achievable Zernike amplitudes for a targeted residual wavefront RMS error of $\lambda/10$ (at 632.8 nm) using the data from (a).

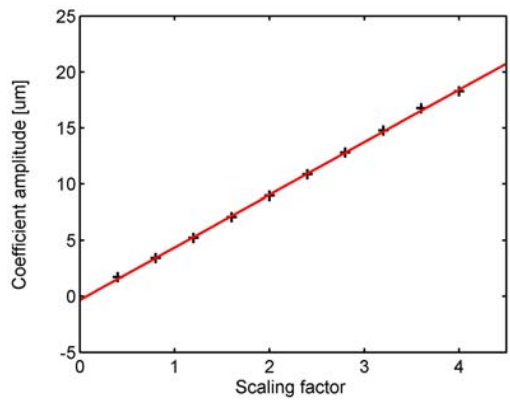

Fig. 10. Amplitude of $Z_2^0$ as a function of a current scaling factor relative to the currents that were found to produce a $Z_2^0$ wavefront having a PV amplitude of 4 µm. A linear fit of the data is shown in red.

The HASO wavefront sensor was used to confirm the results at higher amplitudes. For example, Fig. 10 shows the amplitude of $Z_2^0$ as a function of a current scaling factor relative to the current vector that was found to produce it with at a PV amplitude of 4 μm. The results show that the linearity of the FDM still holds at large amplitudes. As another high amplitude example, Fig. 11 shows an astigmatism term having a PV amplitude of 20 μm (left) and its residual wavefront error at right. The residual wavefront has a RMS of 0.044 μm. Both Fig. 10 and 11 confirm the large wavefront amplitudes the FDM can achieve.

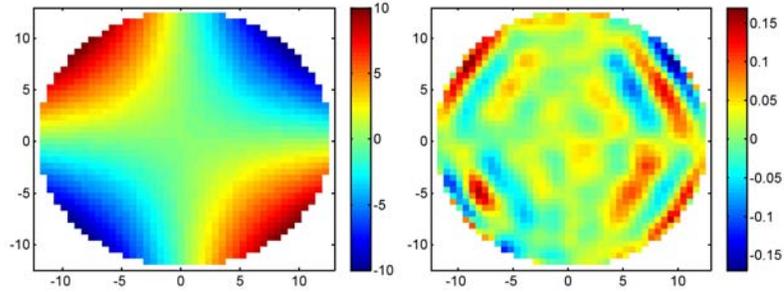

Fig. 11. A 20 μm PV astigmatism produced by the 91-actuator FDM (left). The residual wavefront (right) has a RMS error of 0.044 μm. Grid units are in millimeters and amplitude is measured in μm. Wavefronts measured using an Imagine Optics HASO HR44 wavefront sensor.

*4.2 Combination of Zernikes*

In section 4.1, we obtained current vectors to produce the first 36 Zernike polynomials. Because the system now has a linear response, it should be possible to produce wavefronts, expressed as a combination of these Zernikes, by linearly adding the current vector of each Zernike scaled by the appropriate amount of each coefficient. To test this, we constructed a targeted wavefront by using a combination of Zernikes $Z_2^{-2}$, $Z_2^0$, $Z_2^2$, $Z_3^{-1}$ and $Z_3^1$ having coefficient amplitudes of 0.4, 0.3, 0.3, 0.3 and 0.1 μm respectively. This theoretical wavefront can be seen at the left of Fig. 11.

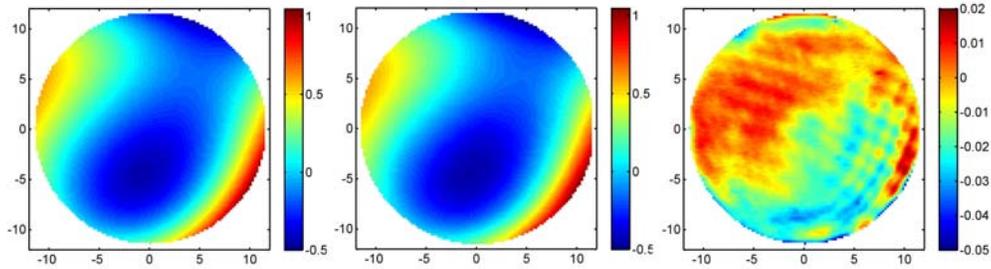

Fig. 12. Wavefronts obtained by combination of Zernikes polynomials. The targeted wavefront can be seen at left while the FDM experimental wavefront is at center. The residual wavefront is shown at right. Residual wavefront RMS error is 0.01 μm. Grid units are in millimeters and amplitude is measured in μm.

The currents obtained for each of the 5 Zernikes were scaled and added according to their coefficient content in the targeted wavefront. This current vector was then supplied to the actuators and the resulting wavefront can be seen at the middle of Fig. 12. The residual wavefront error can be seen at the right of Fig. 8 and the residual wavefront RMS error is 0.01 μm. This demonstrates that this FDM can also be driven in modal mode and also allows for open-loop driving when the Zernike content of the incoming wavefront is known. This could have a major impact in optical null-testing of aspherics for example [17].

*4.3 Repeatability*

We verified the repeatability of the FDM at producing Zernikes of constant amplitude over time using their corresponding current vector. The current vectors of Zernike $Z_2^{-2}$, $Z_2^0$ and $Z_3^{-1}$ were scaled and applied to the actuators four times over the course of three days and the corresponding Zernike coefficient were extracted from the ZYGO measurements. Table 1 presents the results. It can be seen that once an optimized current vector that produces a given Zernike is found, this current vector can be scaled to produce Zernikes (or combination of Zernikes) of varying amplitude, as also demonstrated by Fig. 10, and that this calibration remains constant over time. From the data of Table 1, we computed that the over-time repeatability of the FDM at producing Zernikes having specific targeted coefficient amplitudes is better than 1.5%.

Table 1. Repeatability measurements of the 91-actuator FDM

| Zernike | $Z_2^{-2}$ | $Z_2^0$ | $Z_3^{-1}$ |
| --- | --- | --- | --- |
| Targeted amplitude [µm] | 1.500 | 1.000 | 0.500 |
| *Measurement #1* | 1.504 | 1.012 | 0.500 |
| *Measurement #2* | 1.503 | 0.999 | 0.502 |
| *Measurement #3* | 1.498 | 0.998 | 0.498 |
| *Measurement #4* | 1.514 | 1.012 | 0.504 |

**5 Conclusion**

We demonstrated a 91-actuator magnetic liquid mirror that can be driven by a standard zonal driving technique. The linear response of the mirror is obtained by superposing a uniform magnetic field to that of the actuators. This external constant magnetic field also amplifies the response of the actuators, allowing strokes of tens of microns. A non-linear response was a major inconvenience on previous FDMs. It will be much simpler to implement a closed-loop system now that there exists a way to linearly drive FDMs.

The first 36 Zernikes polynomials were reproduced with a residual wavefront RMS error lower than 0.05 µm for all terms having a targeted PV amplitude of 4 µm. We were limited to this low-amplitude by the Fizeau interferometer which was chosen for its excellent spatial resolution and the lower number of optical components necessary to perform the measurements. A large amplitude wavefront having a PV amplitude of 20 µm was demonstrated using a Shack-Hartmann wavefront sensor. Linearity of the FDM response was demonstrated for a defocus coefficient of 2 to 17 µm. Even larger amplitudes can easily be obtained but we are limited by the maximum slopes the Shack-Hartmann can measure. From Fig. 9 we conclude that the achievable mean PV amplitude over all terms, while keeping a residual error of $\lambda/10$, is over 30 µm with some terms even over 70 µm. Even greater amplitudes and lower residuals could be obtained by increasing the number of actuators and using a higher current for the Maxwell coil. These values are higher than the maximum 50 µm (tilt) of the commercially available Imagine Eyes mirao 52-e deformable mirror that relies on the electromagnetic deformation of a magnetic membrane, which is, technologically, the deformable mirror that is the closest to a FDM.

The 45% coupling constant of the FDM could be somewhat reduced on future versions by increasing the inter-actuator distance. This would help reducing residual errors on wavefronts containing higher spatial frequencies.

No noticeable power dissipation in the actuators was detected within the available +/- 200 mA current range. Power dissipation was only noticed for Maxwell coil currents over 1.5 A. But, from crude measurements, using the Maxwell coil below 1 A already makes possible the production of wavefronts having PV amplitudes of over 50 μm. Construction of the Maxwell coil could also be optimized, e.g. using lower wire gauge to reduce resistance, to give an amplification factor allowing PV wavefront amplitudes over one hundred microns. Ferrofluids have a low vapor pressure and we did not notice any evaporation during the months we have been using the FDM. As for the coatings, chemists within our team are still working on optimizing their reflectivity and robustness.

The optimized current vectors obtained for each Zernike term can be combined linearly to form high amplitude wavefronts having low residuals with over-time repeatability better than 1.5%.

While the early FDMs that we produced [7] seemed promising, they were early-stage prototypes with parameters substantially inferior to that of many commercial deformable mirrors (e.g. smaller number of actuators and non-linear response). However, the 91-actuator FDM presented in this paper has a number of actuators comparable to the number available in high-end commercial mirrors as well as a superior performance both in available stroke and residuals. A remarkable feature of this prototype is that it was made without sophisticated technology and within a short time scale using the limited resources typically available in a University laboratory. We emphasize this point because it illustrates the other major advantages of FDMs: their simplicity and low-cost. These new FDMs that have a linear response can produce wavefront amplitudes that no other commercial deformable mirror can produce. By its low cost, simple design and low residuals, it represents a promising solution for both adaptive optics applications and optical testing.

## Acknowledgments

This research was supported by the Natural Sciences and Engineering Research Council of Canada and the Canadian Institute for Photonic Innovations.